\documentstyle[12pt]{article}

\def\Dot#1{{\kern0.5pt
                   {#1} \kern-5.05pt \raise5.8pt\hbox{$\textstyle.$}\kern
0.5pt}}

\skewchar\fivmi='177
\skewchar\sixmi='177
\skewchar\sevmi='177
\skewchar\egtmi='177
\skewchar\ninmi='177
\skewchar\tenmi='177
\skewchar\elvmi='177
\skewchar\twlmi='177
\skewchar\frtnmi='177
\skewchar\svtnmi='177
\skewchar\twtymi='177
\def\@magscale#1{ scaled \magstep #1}
\skewchar\fivsy='60
\skewchar\sixsy='60
\skewchar\sevsy='60
\skewchar\egtsy='60
\skewchar\ninsy='60
\skewchar\tensy='60
\skewchar\elvsy='60
\skewchar\twlsy='60
\skewchar\frtnsy='60
\skewchar\svtnsy='60
\skewchar\twtysy='60

\catcode`@=11
\def\un#1{\relax\ifmmode\@@underline#1\else
        $\@@underline{\hbox{#1}}$\relax\fi}
\catcode`@=12

\let\du=\d                      

\def\a{\alpha}

\def\d{\delta}
\def\e{\epsilon}

\def\s{\sigma}
\def\t{\tau}

\def\z{\zeta}
\def\D{\Delta}

\def\lin{\vrule width0.5pt height5pt depth1pt}
\def\dpx{{{ =\hskip-3.75pt{\lin}}\hskip3.75pt }}

\def\dslash{\not{\hbox{\kern-2pt $\partial$}}}
\def\Dslash{\not{\hbox{\kern-4pt $D$}}}
\def\pslash{\not{\hbox{\kern-2.3pt $p$}}}
 \newtoks\slashfraction
 \slashfraction={.13}
 \def\slash#1{\setbox0\hbox{$ #1 $}
 \setbox0\hbox to \the\slashfraction\wd0{\hss \box0}/\box0 }
 
\def\Sc#1{{\hbox{\sc #1}}}      
\font\ro=cmsy10                          
\def\kcr{{\hbox{\ro \char'170}}}                
\def\ktl{{\hbox{\ro \char'170}}}        
\def\ktr{{\hbox{\ro \char'170}}}        
\def\kbl{{\hbox{\ro \char'170}}}        
\def\kbr{{\hbox{\ro \char'170}}}        

  \def\pp{{\mathchoice
          {
              \kern 1pt%
              \raise 1pt
              \vbox{\hrule width5pt height0.4pt depth0pt
                    \kern -2pt
                    \hbox{\kern 2.3pt
                          \vrule width0.4pt height6pt depth0pt
                          }
                    \kern -2pt
                    \hrule width5pt height0.4pt depth0pt}%
                    \kern 1pt
           }
            {
              \kern 1pt%
              \raise 1pt
              \vbox{\hrule width4.3pt height0.4pt depth0pt
                    \kern -1.8pt
                    \hbox{\kern 1.95pt
                          \vrule width0.4pt height5.4pt depth0pt
                          }
                    \kern -1.8pt
                    \hrule width4.3pt height0.4pt depth0pt}%
                    \kern 1pt
            }
            {
              \kern 0.5pt%
              \raise 1pt
              \vbox{\hrule width4.0pt height0.3pt depth0pt
                    \kern -1.9pt  
                    \hbox{\kern 1.85pt
                          \vrule width0.3pt height5.7pt depth0pt
                          }
                    \kern -1.9pt
                    \hrule width4.0pt height0.3pt depth0pt}%
                    \kern 0.5pt
            }
            {
              \kern 0.5pt%
              \raise 1pt
              \vbox{\hrule width3.6pt height0.3pt depth0pt
                    \kern -1.5pt
                    \hbox{\kern 1.65pt
                          \vrule width0.3pt height4.5pt depth0pt
                          }
                    \kern -1.5pt
                    \hrule width3.6pt height0.3pt depth0pt}%
                    \kern 0.5pt
            }
        }}

  \def\mm{{\mathchoice
                       {
                             \kern 1pt
               \raise 1pt    \vbox{\hrule width5pt height0.4pt depth0pt
                                  \kern 2pt
                                  \hrule width5pt height0.4pt depth0pt}
                             \kern 1pt}
                       {
                            \kern 1pt
               \raise 1pt \vbox{\hrule width4.3pt height0.4pt depth0pt
                                  \kern 1.8pt
                                  \hrule width4.3pt height0.4pt depth0pt}
                             \kern 1pt}
                       {
                            \kern 0.5pt
               \raise 1pt
                            \vbox{\hrule width4.0pt height0.3pt depth0pt
                                  \kern 1.9pt
                                  \hrule width4.0pt height0.3pt depth0pt}
                            \kern 1pt}
                       {
                           \kern 0.5pt
             \raise 1pt  \vbox{\hrule width3.6pt height0.3pt depth0pt
                                  \kern 1.5pt
                                  \hrule width3.6pt height0.3pt depth0pt}
                           \kern 0.5pt}
                       }}

\def\bo{{\raise.15ex\hbox{\large$\Box$}}}               
\def\Bo{ {}_{{\raise.15ex\hbox{\large$\Box$}}}}         
\def\pa{\partial}                                       
\def\pr{\prod}                                          
\def\TH{{\raise.2ex\hbox{$\displaystyle \bigodot$}\mskip-4.7mu \llap H \;}}
\def\face{{\raise.2ex\hbox{$\displaystyle \bigodot$}\mskip-2.2mu \llap {$\ddot
        \smile$}}}                                      

\def\sp#1{{}^{#1}}                              
\def\leftrightarrowfill{$\mathsurround=0pt \mathord\leftarrow \mkern-6mu
        \cleaders\hbox{$\mkern-2mu \mathord- \mkern-2mu$}\hfill
        \mkern-6mu \mathord\rightarrow$}
\def\dvec#1{\vbox{\ialign{##\crcr
        \leftrightarrowfill\crcr\noalign{\kern-1pt\nointerlineskip}
        $\hfil\displaystyle{#1}\hfil$\crcr}}}           

\def\fracm#1#2{\hbox{\large{${\frac{{#1}}{{#2}}}$}}}
\def\frac#1#2{{\textstyle{#1\over\vphantom2\smash{\raise.20ex
        \hbox{$\scriptstyle{#2}$}}}}}                   
\def\sfrac#1#2{{\vphantom1\smash{\lower.5ex\hbox{\small$#1$}}\over
        \vphantom1\smash{\raise.4ex\hbox{\small$#2$}}}} 
\def\bfrac#1#2{{\vphantom1\smash{\lower.5ex\hbox{$#1$}}\over
        \vphantom1\smash{\raise.3ex\hbox{$#2$}}}}       
\def\afrac#1#2{{\vphantom1\smash{\lower.5ex\hbox{$#1$}}\over#2}}    

\newskip\humongous \humongous=0pt plus 1000pt minus 1000pt
\def\caja{\mathsurround=0pt}
\def\eqalign#1{\,\vcenter{\openup2\jot \caja
        \ialign{\strut \hfil$\displaystyle{##}$&$
        \displaystyle{{}##}$\hfil\crcr#1\crcr}}\,}
\newif\ifdtup

\def\ref#1{$\sp{#1)}$}

\def\np#1{{\it Nucl. Phys.} {\bf B#1}}

\def\cqg#1{{\it Class. Quan. Grav.} {\bf #1}}

\def\prl#1{{\it Phys. Rev. Letts.} {\bf #1}}
\def\pr#1{{\it Phys. Rev.} {\bf #1}}
\def\pl#1{{\it Phys. Letts.} {\bf #1B}}

\parskip=\medskipamount                 
\thicklines                         

\def\oldheadpic{                                
        \setlength{\unitlength}{.4mm}
        \thinlines
        \par
        \begin{picture}(349,16)
        \put(325,16){\line(1,0){4}}
        \put(330,16){\line(1,0){4}}
        \put(340,16){\line(1,0){4}}
        \put(335,0){\line(1,0){4}}
        \put(340,0){\line(1,0){4}}
        \put(345,0){\line(1,0){4}}
        \put(329,0){\line(0,1){16}}
        \put(330,0){\line(0,1){16}}
        \put(339,0){\line(0,1){16}}
        \put(340,0){\line(0,1){16}}
        \put(344,0){\line(0,1){16}}
        \put(345,0){\line(0,1){16}}
        \put(329,16){\oval(8,32)[bl]}
        \put(330,16){\oval(8,32)[br]}
        \put(339,0){\oval(8,32)[tl]}
        \put(345,0){\oval(8,32)[tr]}
        \end{picture}
        \par
        \thicklines
        \vskip.2in}
\def\oldtitle#1#2#3#4{\oldheadpic\begin{center}\vglue.5in{\large\bf #1}\\[.6in]
        {#2}\\[.1in] {\it Department of Physics and Astronomy}\\
        {\it University of Maryland, College Park, MD 20742}\\[.6in]
        Physics Publication \#{#3}\\ {#4}\\[1.5in] {\bf ABSTRACT}\\[.1in]
        \end{center} \begin{quotation}}                 
\def\oldTitle#1#2#3#4#5#6#7{\oldheadpic\begin{center} \vglue .4in
        {\large\bf #1}\\[.4in]
        {#2}\\[.1in] {\it Department of Physics and Astronomy}\\
        {\it University of Maryland, College Park, MD 20742}\\[.1in]
        {#3}\\[.1in] {\it {#4}}\\ {\it {#5}}\\[.4in]
        Physics Publication \#{#6}\\ {#7}\\[.5in] {\bf ABSTRACT}\\[.1in]
        \end{center} \begin{quotation}}                 
\def\border{                                            
        \setlength{\unitlength}{1mm}
        \newcount\xco
        \newcount\yco
        \xco=-21
        \yco=12
        \begin{picture}(140,0)
        \put(\xco,\yco){$\ktl$}
        \advance\yco by-1
        {\loop
        \put(\xco,\yco){$\kcr$}
        \advance\yco by-2
        \ifnum\yco>-240
        \repeat
        \put(\xco,\yco){$\kbl$}}
        \xco=158
        \yco=12
        \put(\xco,\yco){$\ktr$}
        \advance\yco by-1
        {\loop
        \put(\xco,\yco){$\kcr$}
        \advance\yco by-2
        \ifnum\yco>-240
        \repeat
        \put(\xco,\yco){$\kbr$}}
        \put(-20,13){\tiny University of Maryland Elementary Particle
Physics University of Maryland Elementary Particle Physics University of
Maryland Elementary Particle Physics}
        \put(-20,-241.5){\tiny University of Maryland Elementary
Particle Physics University of Maryland Elementary Particle Physics
University of Maryland Elementary Particle Physics}
        \end{picture}
        \par\vskip-8mm}
\def\bordero{                                           
        \setlength{\unitlength}{1mm}
        \newcount\xco
        \newcount\yco
        \xco=-31
        \yco=12
        \begin{picture}(140,0)
        \put(\xco,\yco){$\ktl$}
        \advance\yco by-1
        {\loop
        \put(\xco,\yco){$\kclr}
        \advance\yco by-2
        \ifnum\yco>-240
        \repeat
        \put(\xco,\yco){$\kbl$}}
        \xco=151
        \yco=12
        \put(\xco,\yco){$\ktr$}
        \advance\yco by-1
        {\loop
        \put(\xco,\yco){$\kcr$}
        \advance\yco by-2
        \ifnum\yco>-240
        \repeat
        \put(\xco,\yco){$\kbr$}}
        \put(-20,12){\ooo bacdefghidfghghdhededbihdgdfdfhhdheidhdhebaaahjhhdahba

hgdedge
   hgfdiehhgdigicba}
        \put(-20,-241.5){\ooo ababaighefdbfghgeahgdfgafagihdidihiidhiagfedhadbfd

ecdcdfa
   gdcbhaddhbgfchbgfdacfediacbabab}
        \end{picture}
        \par\vskip-8mm}
\def\headpic{                                           
        \indent
        \setlength{\unitlength}{.4mm}
        \thinlines
        \par
        \begin{picture}(29,16)
        \put(165,16){\line(1,0){4}}
        \put(170,16){\line(1,0){4}}
        \put(180,16){\line(1,0){4}}
        \put(175,0){\line(1,0){4}}
        \put(180,0){\line(1,0){4}}
        \put(185,0){\line(1,0){4}}
        \put(169,0){\line(0,1){16}}
        \put(170,0){\line(0,1){16}}
        \put(179,0){\line(0,1){16}}
        \put(180,0){\line(0,1){16}}
        \put(184,0){\line(0,1){16}}
        \put(185,0){\line(0,1){16}}
        \put(169,16){\oval(8,32)[bl]}
        \put(170,16){\oval(8,32)[br]}
        \put(179,0){\oval(8,32)[tl]}
        \put(185,0){\oval(8,32)[tr]}
        \end{picture}
        \par\vskip-6.5mm
        \thicklines}
\def\title#1#2#3#4{\border\headpic {\hbox to\hsize{#4 \hfill UMDEPP #3}}\par
        \begin{center} \vglue .5in {\large\bf #1}\\[.6in]
        {#2}\\[.1in] {\it Department of Physics and Astronomy}\\
        {\it University of Maryland, College Park, MD 20742}\\[1.5in]
        {\bf ABSTRACT}\\[.1in] \end{center} \begin{quotation}}  
\def\Title#1#2#3#4#5#6#7{\border\headpic
        {\hbox to\hsize{#7 \hfill UMDEPP #6}}\par
        \begin{center} \vglue .4in {\large\bf #1}\\[.4in]
        {#2}\\[.1in] {\it Department of Physics and Astronomy}\\
        {\it University of Maryland, College Park, MD 20742}\\[.1in]
        {#3}\\[.1in] {\it {#4}}\\ {\it {#5}}\\[.5in] {\bf ABSTRACT}\\[.1in]
        \end{center} \begin{quotation}}                 
\def\endtitle{\end{quotation}\newpage}                  

\def\sect#1{\bigskip\medskip \goodbreak \noindent{\bf {#1}} \nobreak \medskip}

\begin{document}

\def\[{\lfloor{\hskip 0.35pt}\!\!\!\lceil}
\def\]{\rfloor{\hskip 0.35pt}\!\!\!\rceil}
\def\asym{({\scriptstyle 1\leftrightarrow \scriptstyle 2})}
\def\Lag{{\cal L}}
\def\du#1#2{_{#1}{}^{#2}}
\def\ud#1#2{^{#1}{}_{#2}}
\def\dud#1#2#3{ {_{#1}{}^{#2}{}}_{#3} }
\def\udu#1#2#3{ {^{#1}{}_{#2}{}}^{#3} }

\def\pl#1#2#3{Phys.~Lett.~{\bf {#1}B} (19{#2}) #3}
\def\np#1#2#3{Nucl.~Phys.~{\bf B{#1}} (19{#2}) #3}
\def\prl#1#2#3{Phys.~Rev.~Lett.~{\bf #1} (19{#2}) #3}
\def\pr#1#2#3{Phys.~Rev.~{\bf D{#1}} (19{#2}) #3}
\def\cqg#1#2#3{Class.~and Quant.~Gr.~{\bf {#1}} (19{#2}) #3} 
\def\ula{{\un a}} 
\def\ulb{{\un b}} 
\def\ulc{{\un c}}
\def\uld{{\un d}}
\def\ulA{{\un A}} 
\def\ulM{{\underline M}} 
\def\cdm{{\Sc D}_{--}}
\def\cdp{{\Sc D}_{++}}
\def\vTheta{\check\Theta}
\def\Kappa{K}
\def\Pisl{{\Pi\!\!\!\! /}}
\def\fracm{\frac}
\def\lin{\vrule width0.5pt height5pt depth1pt}
\def\dpx{{{ =\hskip-3.75pt{\lin}}\hskip3.75pt }}

\border\headpic {\hbox to\hsize{June 1998 \hfill UMDEPP 98-114}}\par
\begin{center}
\vglue .4in
{\large\bf Superspace\\
Geometrical Representations of\\
Extended Super Virasoro Algebras\footnote{Research 
supported in part by by NSF grant \# PHY-96-43219.} }\\[.2in]
S. James Gates, Jr. and Lubna Rana \\[.1in]
{\it Department of Physics\\
University of Maryland at College Park\\
College Park, MD 20742-4111, USA} \\[.5in]

{\bf ABSTRACT}\\[.1in]
\end{center}
\begin{quotation}

Utilizing sets of super-vector fields (derivations), explicit expressions are obtained for; (a.) the 1D, $N$-extended 
superconformal algebra, (b.) the 1D, $N$-extended super Virasoro algebra 
for $N \, = \, 1,\, 2,$ and $4$ and (c.) a geometrical realization (${\cal G R}$) 
covering algebra that contains the super Virasoro algebra for arbitrary 
values of $N$. By use of such vector fields, the super Virasoro algebra is 
embedded as a geometrical and model-independent structure in 1D and 
2D $\aleph_0$-extended superspace.

\endtitle

\sect{I. Introduction}

Spacetime symmetry groups possess many distinct representations. 
Consequently this is so for their supersymmetric extensions. Moreover 
when considering theories defined over extended manifolds, such symmetry 
groups continue to play important roles. In the realm of superstring 
theory, the role of the superconformal group is taken over by the 
super-Virasoro group \cite{VA,NSR}. For spacetime symmetry groups, among 
the simplest faithful representations are those 
constructed from ``derivations'' or ``vector fields'' where the symmetry 
generators are represented in terms of functions of the coordinates of 
the manifold which multiply linear derivatives with respect to those 
coordinates.  The set of all such derivations for a given algebra is
sometimes referred to as a ``D-module.''

The simplest of super manifolds are those associated with
superspaces that possess a single bosonic dimension. For some time
\cite{GRn} we have been engaged in a study of 1D superspaces and the 
superfield representations that can be defined over such manifolds.  
One previous useful result found was the complete off-shell 
formulation of NSR spinning particle models for arbitrary numbers of 
extended supersymmetries on the worldline \cite{GRn3}. Although there 
are a number of features that remain to be explored in our previous 
work, it should be apparent that there is much to learn by studying 
such systems in great detail. A spectacular (but not obviously related) 
example of the utility of 1D supersymmetric models has been provided by 
the proposal that a microscopic description of M-theory is as a matrix 
theory of a 1D supersymmetric system \cite{Mthy}.  

Thus, a primary purpose of this letter is to present linear differential 
operators to represent the super Virasoro algebra for all values of $N$, 
the degree of supersymmetric extension.  Such constructions are intrinsically 
geometrical or kinematical having no {\it {a}} {\it {priori}} relationships 
to properties of any given model and provide a basis for the
study of super Virasoro algebras that is independent of the dynamical
system in which they are realized.

\sect{II. 1D Superconformal Derivations}

In the following, $P$, $K$, $Q_{\rm I}$, $S_{\rm I}$, $\D$ and $T_{\rm {I 
\, J}}$ are the momentum, special conformal, supersymmetry,
s-supersymmetry, dilatation and SO($N$) generators. It is a simple exercise 
to find a set of derivations with respect to a one dimensional superspace
with coordinates $\z^{\rm I}$ and $\t$ in terms of which these abstract
operators may be realized
$$\eqalign{
P \, &=~  i \, \pa_{\t} ~~~,  ~~~ {~~~~~~~~~~~~~~~~~~~}
K \, ~=~ i \, \left( \t^2 \, \pa_\t \, + \, \t \, \zeta^{\rm I} \, \pa_{
\rm I} \right) ~~~, \cr
Q_{\rm I} \, &=~ i \, \left( \pa_{\rm I} \, -\, 2i\, \zeta_{\rm I} \, 
\pa_\t \right) ~~~, ~~~
~~\,S_{\rm I} \, ~=~ i \, \t \, \pa_{\rm I} \, + \, 2\, \t \, \zeta_{\rm I} 
\, \pa_{\t} \, + \, 2 \, \zeta_{\rm I} \, \zeta^{\rm J} \, \pa_{\rm J} 
~~~, \cr
\D \, &=~ i \, \left( \t \, \pa_{\t} \, + \, \frac{1}{2} \zeta^{\rm I} 
\, \pa_{\rm I} \right) ~~~, ~~~ T_{\rm {I \, J}} \, ~=~ i \, \left( 
\zeta_{\rm I} \, \pa_{\rm J} \, - \, \zeta_{\rm J}  \, \pa_{\rm I}
\right) ~~~. 
} \eqno(1) $$ 

\noindent The generic 1D, $N$-extended superconformal algebra is given by:
$$\eqalign{ {~~~~~~~}
\left[ \D \, , \, P \right\} ~~=~~~& -i \, P ~~,~~ \left[ \D \, , \, 
Q_{\rm I} \right\} ~=~ - i \fracm 12  Q_{\rm I} ~~,~~ \left[ K \, , \, 
Q_{\rm I} \right\} ~~=~~ - \, i \, S_{\rm I} ~~~, {~~~~}\cr
\left[ \D \, , \, K \right\} ~~=~~~& i \, K ~~, ~~ \left[ \D \, , \, 
S_{\rm I} \right\} ~=~ i \fracm 12  S_{\rm I} ~~,~~ \left[ P \, , \, 
S_{\rm I} \right\} ~~=~~ i \, Q_{\rm I} ~~~, \cr
\left[ Q_{\rm I} \, , \, Q_{\rm J}  \right\}  ~~=~~~& 4 \, \d_{\rm {I 
\, J}} \, P ~~,~~~ \left[ S_{\rm I} \, , \, S_{\rm J} \right\}  
~~=~~ 4 \,  \d_{\rm {I \, J}} \, K ~~ ~~~,\cr 
\left[ P \, , \, K \right\} ~~=~~~& - \, i 2 \, \D ~~,~~ 
\left[ Q_{\rm I} \, , \, S_{\rm J}  \right\}  ~~=~~ 4 \, 
\d_{\rm {I \, J}} \ \D \, + \, 2 \, T_{\rm {I \, J}}  ~~,~~ \cr 
\left[ T_{\rm {I \, J}} \, , \, Q_{\rm K} \right\} ~~=~~~& - i 
\d_{\rm {I \, K}} \, Q_{\rm J} \, + \, i \, \d_{\rm {J \, K}} \, 
Q_{\rm I} ~~,~~ \cr 
\left[ T_{\rm {I \, J}} \, , \, S_{\rm K} \right\} ~~=~~~& - i 
\d_{\rm {I \, K}} \, S_{\rm J} \, + \, i \, \d_{\rm {J \, K}} \, 
S_{\rm I} ~~,~~ \cr
\left[ T_{\rm {I \, J}} \, , T_{KL} \right\} ~~=~~~& i \d_{\rm 
{J \, K}} \, T_{\rm {I \, L}} \, - \, i \d_{\rm {J \,L}} \, T_{\rm {I 
\, K}} \, + \, i \d_{\rm {I \, L}} \, T_{\rm {J \, K}} \, - \, i \d_{\rm 
{I \, K}} \, T_{\rm {J\, L}}~~~.
 } \eqno(2) $$

Given these operators, they also define the 2D superconformal 
algebras. For example, the generators of the 2D, 
($N$,0) superconformal algebras are constructed by replacing the 
coordinates as $\t ~\to~ \s^{\pp}$, $\z^{\rm I} ~\to~ \z^{+ \, 
{\rm I}}$ in the derivations in (1) and simultaneously replacing 
the 1D generators by 2D generators via,
$$ 
P ~\to~ P_{\pp} ~,~~ K ~\to~ K_{\mm} ~,~~ \D ~\to~ \fracm 12 (\,
\D \, + \, L \,) ~,~~ Q_{\rm I} ~\to~ Q_{+ \, \rm I} ~,~~ S_{\rm I} 
~\to~ S_{- \, \rm I} ~~~. 
\eqno(3) $$
We must also add the generators that are missing from the
above ``oxidation'' up to two dimensions.  For this purpose we
replace the coordinates and derivatives in (1) according to 
$\t ~\to~ \s^{\mm}$, $\z^{\rm I} ~\to~ 0$ and $\pa_{\rm I} ~\to~ 0$
and also simultaneously replace the 1D generators by 2D generators
via,
$$ 
P ~\to~ P_{\mm} ~,~~ K ~\to~ K_{\pp} ~,~~ \D ~\to~ \fracm 12 (\,
\D \, - \, L \,) ~,~~ Q_{\rm I} ~\to~ 0 ~,~~ S_{\rm I} 
~\to~ 0 ~~~. 
\eqno(4) $$
The quantity denoted by $L$ above is the generator of 2D Lorentz rotations 
which is obviously absent in 1D. By an obvious modification of the above 
described replacement procedure, the generators of all 2D, (p, q) 
superconformal algebras may be obtained.

This approach also yields a new viewpoint on why 1D and 2D theories
are singled out as special when considered from this geometrically
based construction. The 1D theory may be regarded as the fundamental 
representation. The oxidation argument above shows that the 2D theory
is constructed essentially as two completely independent copies 
of the 1D theory. For no dimension D $\ge$ 3 does the conformal
group ``factorize'' into independent copies of the 1D group.

Another useful aspect of this representation is that it permits us to see 
that the case of $N \, =\, 4$ is exceptional and makes clear the geometrical 
origin of the ``small'' versus ``large'' $N \,= \,4$ superconformal algebras.  
We observe that in the case of $N \,=\, 4$, there exists a Levi-Civita 
tensor $\e_{\rm {I \, J \, K \, L}}$ which may be used to ``deform'' 
some of the derivations according to
$$
\eqalign{ {~~~~}
S_{\rm I} (\ell)  \, &\equiv~ i \, \t \, \pa_{\rm I} \, + \, 2\, \t \, \zeta_{
\rm I} \, \pa_{\t} \, + \, 2 \, \zeta_{\rm I} \, \zeta^{\rm J} \, \pa_{
\rm J}  \,+\,  \ell \,  (\, \e_{\rm {I \, J \, K \, L}} \,\zeta^{
\rm J} \zeta^{\rm K} \pa^{\rm L} \, - \, i \, 4 \, \zeta^{(3)}{}_{\rm I} 
 \pa_{\t} \, ) ~~~, \cr
K (\ell) \, &\equiv~ i \, \left[~ \t^2 \, \pa_\t \, + \, \t \, \zeta^{\rm I} \, 
\pa_{\rm I} ~-~ i 2 \, \ell\, ( \, \zeta^{(3) \, \rm I} \, \pa_{\rm I} ~+~ 
i \, 4\, \zeta^{(4)} \,  \pa_{\t} \, ) ~ \right]~~~, \cr
T_{\rm {I \, J}} (\ell)  \, &\equiv~ i \, \left[~ \zeta_{\rm I} \, \pa_{\rm J} \, 
- \, \zeta_{\rm J} \, \pa_{\rm I} \,-\,  \ell \, \e_{\rm {I \, J \, K \, L}} 
\, \zeta_{\rm K} \, \pa_{\rm L} ~ \right] ~~~.
} \eqno(5) $$
Above the quantities $\zeta^{(3) \, \rm I}$ and $\zeta^{(4)}$ are defined
by
$$
\, \zeta_{\rm I} \, \zeta_{\rm J} \, \zeta_{\rm K} ~\equiv~  \e_{\rm {I \, J 
\, K \, L}} \, \zeta^{(3) \, \rm L} ~~~,~~~  \zeta_{\rm I} \, \zeta_{\rm J} \, 
\zeta_{\rm K}  \, \zeta_{\rm L} ~\equiv~  \e_{\rm {I \, J \, K \, L}} \, 
\zeta^{(4)} ~~. 
\eqno(6) $$
The remaining generators of the deformed algebra remain unchanged
from their definitions in (1).  The exceptional role of the $N \,=\,4$ theory is 
also apparent from the point of view of dimensional analysis. It is only for 
the value of $N \,=\, 4$ that terms of a proper dimension (-1/2 for $S_{\rm I}$, 
0 for $T_{\rm {I\, J}}$ and $-1$ for $K$) exist that can be used to modify 
some of the generators in an appropriate manner.

The first ten results in the algebra of (2) remain unchanged for the 
$\ell$-deformed algebra. However, for the last three results
we obtain
$$\eqalign{ {~~}
\left[ T_{\rm {I \, J}} \, , \, Q_{\rm K} \right\} &=~ - i 
\d_{\rm {I \, K}} \, Q_{\rm J} \, + \, i \, \d_{\rm {J \, K}} \, 
Q_{\rm I} \, + \, i \ell \, \e_{\rm {I \, J \, K \, L}} Q_{\rm L}
~~,~~ \cr 
\left[ T_{\rm {I \, J}} \, , \, S_{\rm K} \right\} &=~ - i 
\d_{\rm {I \, K}} \, S_{\rm J} \, + \, i \, \d_{\rm {J \, K}} \, 
S_{\rm I} \, + \, i \ell \, \e_{\rm {I \, J \, K \, L}} S_{\rm L}
~~,~~ \cr
\left[ T_{\rm {I \, J}} \, , T_{KL} \right\} &=~ \fracm 12 (
\ell^2 \, + \, 3 ) \, \Big[ ~i \d_{\rm 
{J \, K}} \, T_{\rm {I \, L}} \, - \, i \d_{\rm {J \,L}} \, T_{\rm {J 
\, K}} \, + \, i \d_{\rm {I \, L}} \, T_{\rm {J \, K}} \, - \, i \d_{\rm 
{I \, K}} \, T_{\rm {J\, L}}~ \Big] \cr
&{~~~} + \, \fracm 12 (
\ell^2 \, - \, 1 ) \, \Big[ ~i \d_{\rm {J \, K}} \, {\cal Y}_{\rm {I \, L}} 
\, - \, i \d_{\rm {J \,L}} \, {\cal Y}_{\rm {J \, K}} \, + \, i \d_{\rm 
{I \, L}} \, {\cal Y}_{\rm {J \, K}} \, - \, i \d_{\rm 
{I \, K}} \, {\cal Y}_{\rm {J\, L}}~ \Big]
 ~~~,
 } \eqno(7) $$
where the operator ${\cal Y}_{\rm {I \, J}}$ is defined by
$$
{\cal Y}_{\rm {I \, J}}\, ~\equiv~ i \, \left[~ \zeta_{\rm I} \, 
\pa_{\rm J} \, - \, \zeta_{\rm J} \, \pa_{\rm I} \,+\, \ell \, 
\e_{\rm {I \, J \, K \, L}} \, \zeta_{\rm K} \, \pa_{\rm L} ~ 
\right] ~~~.
\eqno(8) $$

At the special values $\ell = \pm 1$, ${\cal Y}_{\rm {I \, J}}$ does 
not appear in the $\ell$-modified superconformal $N \,=\,4$ algebra. 
This has a dramatic effect. For all values $N \,\ne$ 4, the $T_{\rm {I 
\, J}}$ generators form a representation of O($N$).  For the $N \,=\,4$
case, the deformed $T_{\rm {I \, J}}$ generators form a representation 
of SU(2) if and only if $\ell = \pm 1$.

The appearance of this exceptional 1D, $N \,=\, 4$ theory may also be
related to the appearance of $N \,=\,1$ supersymmetry in four dimensional
spacetime. A 4D spinor is equivalent to a 1D, $N \,=\, 4$ spinor. It is a 
simple matter to show that the $K$-generator of a four dimensional
supersymmetric theory has the form of the $K$-generator in (5) with
$\ell \ne 0$. We thus suspect this plays some role as the 1D worldline
origin of 4D, $N \,=\, 1$ target space supersymmetry.

\sect{III. 1D, $N \,=\, 1$  Superspace and a Super D-module}

If we set all Grassmann coordinates and their derivatives to zero, we are 
left with the bosonic $N \,=\,0$ theory.  It is well known that $P$, $\D$ 
and $K$ form a sub-group of the larger Virasoro algebra. We can define 
the generators
$$
L_{m} ~\equiv~ - \, \t^{m + 1} \pa_{\t} ~~\to ~~
L_{-1} ~\propto~ P ~~,~~ L_{0} ~\propto~ \D ~~,~~ L_{1} ~\propto~ K
  ~~~, 
\eqno(9) $$
and for general values of $m$ we see
$$
[\,  L_{m} ~,~  L_{n} \,\} ~=~ ( \, m \, - \, n ) \,  L_{m \, + \, n}
~~~,
\eqno(10) $$
the characteristic form of the Virasoro commutator algebra.  
The obvious question to ask is, ``What are the explicit expressions 
for $L_n {}^I$, $F_n {}^I$ and $G_r {}^I$ as derivations to represent
the super Virasoro algebra?'' 

We will approach this question by first considering the case of $N \,=\,1$. 
Here we define
$$
\eqalign{ {~~~}
L_{m} &\equiv~ -\, \Big[ \, \t^{m + 1} \pa_{\t} ~+~ \fracm 12
(m \, + \, 1) \, \t^{m} \zeta \, \pa_{\zeta} ~ \Big]  ~~~,  \cr
{~~} F_{m}  ~\equiv~ i \, \t^{{m + \fracm 12} } \, \Big[\, a_0 \, 
\pa_{\zeta} 
&-~ i \, b_0 \, \zeta \, \pa_{\t}  ~ \Big]~~~,  ~~~
G_r  ~\equiv~ i \, \t^{{r + \fracm 12} } \, \Big[\, \, \pa_{\zeta}
~-~ i \, 2 \, \zeta \, \pa_{\t}  ~ \Big] ~~~.
} \eqno(11) 
$$
The reader will note $F_{m}$ and $G_r$ almost possess the same form.
The conditions
$$
G_{- \frac 12} ~ \propto ~ Q ~~~,~~~ G_{\frac 12} ~ \propto ~ S
~~~, 
\eqno(12) $$
have been used to determine the would-be free constants appearing in 
$G_{r}$. We can make $F_m$ identical in form to $G_r$ by picking $a_0 
= 1$ and $b_0 = 2$.  Since for this choice the two operators have 
identical forms, it is useful to clearly state why they are fundamentally 
different\footnote{The choice of $a_0 = 1, \,b_0 = 2$ is also motivated 
by string theory, where these two operators \newline ${~~~~~}$ locally 
satisfy the same algebra.}.  In $F_m$ (the Ramond generator) since $m 
\in Z $ its pre-factor $\t^{m + \frac 12}$ always contains a 
square root of $\t$.  On the other hand in $G_r$ (the Neveu-Schwarz 
generator) since $r \in Z + \fracm 12$, its pre-factor $\t^{r + \frac 12}$ 
never contains a square root of $\t$.  So while the two Grassmann 
generators can possess the same form (by a choice of $a_0$ and $b_0$) 
this appearance is deceptive.  Furthermore, since the index $m$ on the 
$L_m$-operators is integer valued, the choices of whether the 
pre-factor contains a square-root or not are the only allowed values.

It is a matter of direct calculation to show 
$$
\eqalign{ {~~}
[ ~ F_m \, , \, F_n ~\} \, &=~ -\, i \, 2 \, a_0 \, b_0 \, 
L_{m + n} ~~~,~~~
 [ ~ G_r \, , \, G_s ~\} \, ~=~ - \, i \, 4 \, L_{r + s} 
~~~,~~~ \cr
[ ~ L_m \, , \, F_n ~\}\,  &=~ ( \, \fracm 12 m \, - \, n \, 
) \, F_{m + n} ~~~, ~~~ [ ~ L_m \, , \, G_r ~\}\,  ~=~ ( \, \fracm 12 
m \, - \, r \, ) \, G_{m + r} ~~~, ~~~ \cr
[~ L_m \, , \, L_n ~\} \,  &=~ ( \, m \, - \, n \, ) \, L_{m + n} 
~~~. }  \eqno(13) $$
Let us call this the $N \,=\,1$ super-Virasoro algebra. Only the first 
equation in (13) depends explicitly on the constants $a_0$ and 
$b_0$.  Both of these must be nonzero, however.  With these results we 
have constructed a realization of the classical (i.e. un-quantized) 
version of the $N \,=\,1$ super-Virasoro algebra that is implicitly carried 
by the coordinates of 1D, $N \,=\,1$ superspace.

However, having written the usual super-Virasoro generators as 
derivations, we can also calculate $[ F_{m}, G_r \}$. We find
$$ 
\eqalign{ 
[ ~ F_m \, , \, G_r ~\} \, &=~ i \left( 2 a_0 + b_0 \right) 
\Big\{ \, \t^{m + r + 1} \, \pa_{\t} \, + \,  i \left[ \, 
\fracm 12 \,+ \,{( 2 a_0 \,m  \, + \,   b_0 \,r \,) \over ( 2 a_0 
\, +\, b_0)} \, \right] \, \t^{m + r} \, \zeta \, \pa_{\zeta} 
~\Big\} ~~~. } \eqno(14) 
$$
Clearly, for the  choice $a_0 = 1$ and $b_0 = 2$ we obtain
$$
[ ~ F_m \, , \, G_r ~\} \, ~=~ - \, i \, 4 \,  H_{m+r}~~~, ~~~
H_r ~\equiv~ - \,  \Big[ \, \t^{r + 1} \pa_{\t} ~+~ \fracm 12
(m \, + \, 1) \, \t^{r} \zeta \, \pa_{\zeta} ~ \Big]  ~~~.
\eqno(15) $$
This new bosonic operator $H_r$ is related to $L_m$ in the same 
way that $G_r$ is related to $F_m$. We can use this to write things 
using a slightly different notation. We define
$L_{\cal A} \, \equiv \, ( L_m, \, H_r)$ and $G_{\cal A} \, \equiv \, 
( F_m, \, G_r)$. The commutation algebra in this new notation takes 
the form
$$
\eqalign{
[~ L_{\cal A} \, , \, L_{\cal B} ~\} \,  &=~ ( \, {\cal A} \, 
- \, {\cal B} \, ) \, L_{{\cal A} + {\cal B}}
~~~,~~~ [ ~ G_{\cal A} \, , \, G_{\cal B} ~\} \, ~=~ - \, i \, 4 \,  
L_{{\cal A} + {\cal B}}~~~,  \cr
[ ~ L_{\cal A} \, , \, G_{\cal B} ~\} \,  &=~ ( \, \fracm 12 {\cal A}
 \, - \, {\cal B} \, ) \, G_{{\cal A} + {\cal B}} 
~~~. } \eqno(16) $$
The set $\{ L_{\cal A} ,~  G_{\cal A} \}$ is closed under graded 
commutation.  Above we have used the notation ${\cal A}$ to denote an 
index that may be valued in either $Z$ or $Z + \fracm 12$. From these 
commutation relationships it can be seen that $H_s$ is a primary 
(in the language of conformal field theory) operator that ``rotates'' 
the Neveu-Schwarz generator into the Ramond generator and vice-versa. 
This suggests that we identify this operator with a corresponding one 
known in conformal field theory \cite{SS}.  Namely, the operator $H_s$ 
seems similar to the purely chiral part of the generator of spectral 
flows which connects NS and R algebras. Although in the conformal field 
theoretical formulation, such an operator appeared only associated with 
the $N \,=\,2$ theories.

The lesson is clear. If we represent the super Virasoro generators as 
derivations with respect to the coordinates of the simplest real 1D 
superspace, the minimal set which provides a representation of the 
super Virasoro algebra contains all of the generators  in the set
$\{ L_{\cal A} ,~  G_{\cal A} \}$.  We will call this super-Lie algebra 
the 1D, $N \,=\,1$ ``super ${\cal G}{\cal R}$-Virasoro'' algebra 
(where ${\cal G}{\cal R}$ stands for ``geometrical realization'').

Having seen this occurrence for the case of $N \,=\,1$, it is natural 
to ask if this may be extended to higher values of $N$. In fact, the 
$N \,=\,1$, 2 cases have been discussed in a textbook by West 
previously \cite{wst}. The important point that was overlooked in 
his presentation, however, is that if the set of generators $\{ L_m ,~ 
F_m^{\rm I} ,~ G_r^{\rm I} \}$ are represented as derivations, then 
it is {\it {not}} a closed set.  Of course, in string theory customarily 
one does not {\it {simultaneously}} discuss NS and R sector generators.

\sect{IV. New 1D, $N \,=\,1$ Super D-module Operators}

In the last section, we saw that the simplest geometrical realization
of the super-Virasoro algebra leads to an enlarged algebra. However, our
method of construction also points to the existence of another derivation
that is not contained in the super-Virasoro algebra. If we return to 
(11) and choose $a_0 = - i, \, b_0 = i 2$, this defines a new derivation
via
$${\cal D}_r ~\equiv~  \, \t^{{r + \fracm 12} } \, \Big[\, \pa_{\zeta}
~+~ i \, 2 \, \zeta \, \pa_{\t}  ~ \Big]~~~ \to ~~~
 [ ~ {\cal D}_r \, , \, {\cal D}_s ~\} ~=~ - \, i \, 4 \, L_{r + s} ~~~,
\eqno(17) $$
so that
$$
[ ~ {\cal D}_r \, , \, G_s ~\} \, ~=~ i \, 2 \, (r - s) \, {\rm d}_{r+s}~~~, ~~~
{\rm d}_m ~\equiv~ \t^{m} \, \zeta \, \pa_{\zeta} ~~~.
\eqno(18) $$

Before we calculate its graded commutators with $L_{m}$, $F_{m}$,  $G_{r}$, 
and $H_r$, let us note that setting $r = - 1/2$, 
$$
{\cal D}_{- \fracm 12 } ~=\, \, D_{\zeta} ~~~,  
\eqno(19) $$
where $D_{\zeta}$ is the world line supersymmetry covariant derivative. 

The commutation of ${\rm d}_{\cal A}$ with $L_{\cal A}$, ${\cal D}_{\cal A}$ 
and $G_{\cal A}$ yields:
$$
\eqalign{
[ ~L_{\cal A} \, , \, {\rm d}_{\cal B} \, \}  ~=~ -{\cal B} \, {\rm d}_{{\cal A} + 
{\cal B}} ~~~, ~~~ [ ~G_{\cal A} \, , \, {\rm d}_{\cal B} \, \}  ~=~ G_{{\cal 
A} + {\cal B}} ~~~, ~~~ [ ~{\cal D}_{\cal A} \, , \, {\rm d}_{\cal B} \, \}  
~=~ {\cal D}_{{\cal A} + {\cal B}} ~~~,}
\eqno(20) $$
so that the set $\{ L_{\cal A} ,~ G_{\cal A} ,~ {\cal D}_{\cal A} , ~ 
{\rm d}_{\cal A} \}$ is closed under graded commutation. This complete 
set of generators is what we call the ``covering algebra.'' It is larger 
than the super Virasoro algebra which it contains as a proper subgroup. The 
usefulness of the covering algebra can be seen by recalling a past example 
of such a structure.  Within the context of 4D, $N \,=\,1$ superspace, 
Sokatchev \cite{Sok} used precisely a covering algebra to provide
the first definitions of the irreducible superfield projection operators. 
The interpretation of this new infinite set of generators is pretty
clear. These provide the infinite dimensional extension to the superspace 
covariant derivative just as the the usual super Virasoro generators provide 
an infinite dimensional extension of the superconformal generators.

\sect{V. 1D, $\aleph_0$-extended Superspace and the Super Covering 
${\cal G}{\cal R}$ D-module }

~~~~We have previously introduced the notion of 1D, $\aleph_0$-extended 
superspace \cite{GRn2} which is the limit of the union of all possible 
1D superspaces with a finite number of supersymmetries N as this parameter 
approaches infinity. The ideas we have discussed so far in this letter 
permit additional structures to be embedded in $\aleph_0$-extended superspace.  
We introduce the set of derivations (by taking linear combinations of 
these, other bases are also possible as we shall see in the 
exceptional $N \,=\,4$ case)
$$
\eqalign{ {~~~~}
G_{\cal A} {}^{\rm I} &\equiv~ i \, \t^{{\cal A} + \fracm 12}  \, \Big[\, 
\, \pa^{\rm I} ~-~ i \, 2 \, \zeta^{\rm I} \, \pa_{\t} ~ \Big] ~+~ 2 (\, 
{\cal A} \,+\, \fracm 12 \,) \t^{{\cal A} - \fracm 12}  \z^{\rm I} \z^{\rm 
K} \, \pa_{\rm K} ~~~, \cr
L_{\cal A} &\equiv~ -\, \Big[ \, \t^{{\cal A} + 1} \pa_{\t} ~+~ \fracm 12
({\cal A} \, + \, 1) \, \t^{\cal A} \zeta^{\rm I} \, \pa_{\rm I} ~ \Big]  
~~~, \cr
T_{\cal A}^{\, \rm {I \, J }} &\equiv~ \t^{\cal A} \, \Big[ ~ \z^{\rm I} 
\, \pa^{\rm J} ~-~ \z^{\rm J} \, \pa^{\rm I}  ~\Big] ~~~, ~~~
{\rm d}_{\cal A}^{\, I \, J } ~\equiv~ \t^{\cal A} \, \Big[ ~ \z^{\rm I} \, 
\pa^{\rm J} ~+~ \z^{\rm J} \, \pa^{\rm I}  ~\Big] ~~~, \cr
{\cal D}_{\cal A} {}^{\rm I} &\equiv~  \, \t^{{\cal A} + \fracm 12}  \, 
\Big[\, \pa^{\rm I}  ~+~ i \, 2 \, \zeta^{\rm I} \, \pa_{\t}  ~ \Big]  ~+~ 
i \, 2 (\, {\cal A} \, + \, \fracm 12 \,) \t^{{\cal A} - \fracm 12}  \z^{\rm 
I} \z^{\rm K} \, \pa_{\rm K} ~~~,  \cr
R_{\cal A}^{{\rm I}_1 \cdots {\rm I}_p} &\equiv~  (i)^{ [\fracm p2 
]} \, \t^{({\cal A} - \fracm {(p - 2)}2 )} \zeta^{{
\rm I}_1} \cdots \, \zeta^{{\rm I}_p} \, \pa_{\t} ~~~~~~,~~~ p \, = \, 2 , 
\, \dots \, , \, N ~~~, \cr
U_{\cal A}^{{\rm I}_1 \cdots {\rm I}_q} &\equiv~ i \, (i)^{ [\fracm q2 ]} 
\, \t^{({\cal A} - \fracm {(q - 2)}2 )} \zeta^{{\rm 
I}_1} \cdots \, \zeta^{{\rm I}_{q-1}} \, \pa^{{\rm I}_q} ~~~,~~~ q \, = \, 
3 , \, \dots \, , \, N \, + \, 1 ~~~, 
} \eqno(21) 
$$
where $N$ is an arbitrary integer. This set of vector fields is closed 
under graded commutation and as well contains the super Virasoro-like 
sub-algebra for all values of $N$. We shall call this algebra the super 
``${\cal {GR}}$ covering algebra.'' Additionally, the notation in the 
exponents of the factors of $i$ includes the ``greatest integer 
in'' function. So $[\fracm p2 ]$ denotes the greatest integer in 
$\fracm p2 $, etc.  All of these derivations possess engineering 
dimensions (not to be confused with ``scale weight'') of $({\rm {mass
}})^{- \cal A}$ power.

These definitions do not depend on a specific value of $N$ and are thus 
appropriate for the entire 1D, $\aleph_0$ superspace that we  have 
introduced before. Of course, for low values of $N$, not all of the 
generators appear. For example, $T_{\cal A}^{\, \rm {I \, J }}$ and 
$R_{\cal A}^{{\rm I}_1 \cdots {\rm I}_p}$ only appear for superspaces 
with $N$ $\ge$ 2. Generically, $U_{\cal A}^{{\rm I}_1 \cdots {\rm I}_q}$ 
only appears for superspaces with $N$ $\ge$ 3. This set of derivations 
forms a closed algebra under the action of the graded commutator. 
Explicitly for some of the graded commutators we find
$$
\eqalign{
[~ L_{\cal A} \, , \, L_{\cal B} ~\} \,  &=~ ( \, {\cal A} \, 
- \, {\cal B} \, ) \, L_{{\cal A} + {\cal B}}
~~,~~ [ ~ L_{\cal A} \,, \, U_{\cal B}^{{\rm I}_1 \cdots {\rm I}_m}
~ \} ~=~ - \, [ ~ {\cal B} \, + \, \fracm 12 \, m \, {\cal A}
 ~] \, U_{\cal A + \cal B}^{{\rm I}_1 \cdots {\rm I}_m} ~~~,
\cr
[ ~ G_{\cal A}{}^{\rm I} \, , \, G_{\cal B}{}^{\rm J} ~\} \, &=~ 
-\, i \, 4 \,  \d^{{\rm {I\, J}}} L_{{\cal A} + {\cal B}}
~-~ i 2 ({\cal A} - {\cal B} ) \, [ \, \, T_{{\cal A} + {\cal B}
}^{\rm {I\, J}} ~+~ 2 ({\cal A} + {\cal B} ) \, U_{{\cal A} + {\cal 
B}}^{\rm {I\, J \,K}}{}_{\rm K} \,\, ] ~~~,  \cr
[ ~ L_{\cal A} \, , \, G_{\cal B}{}^{\rm I} ~\} \,  &=~ ( \, 
\fracm 12 {\cal A} \, - \, {\cal B} \, ) \, G_{{\cal A} + {\cal 
B}}{}^{\rm I} ~~~,~~~ [ ~ L_{\cal A} \, , \, {\cal D}_{\cal B}
{}^{\rm I} ~\} \,  ~=~ ( \, \fracm 12 {\cal A} \, - \, {\cal B} 
\, ) \, {\cal D}_{{\cal A} + {\cal B}}{}^{\rm I} ~~~,
\cr
[ ~ {\cal D}_{\cal A}{}^{\rm I} \, , \, {\cal D}_{\cal B}{}^{\rm 
J} ~\} \, &=~ -\, i \, 4 \,  \d^{{\rm {I\, J}}} L_{{\cal A} + {\cal 
B}} ~-~ i 2 ({\cal A} - {\cal B} ) \, [ \, \, T_{{\cal A} + {\cal 
B}}^{\rm {I\, J}} ~-~ 2 ({\cal A} + {\cal B} ) \, U_{{\cal A} + {\cal 
B}}^{\rm {I\, J \, K}}{}_{\rm K} \,\, ]  ~~~,  \cr
[ ~ {\cal D}_{\cal A}{}^{\rm I} \, , \, G_{\cal B}{}^{\rm J} ~\} \, 
&=~ i 2 ( {\cal A} - {\cal B}) \, \Big\{ \,  {\rm d}_{\cal A +
{\cal B}}{}^{\rm {I \, J}} \, - \, \d^{\rm {I \, J}}
{\rm d}_{\cal A + {\cal B}}{}^{\rm K}{}_{\rm K} ~+~2 ( {\cal A} + {\cal B})
\, U^{\rm {I\, J \, K}}_{{\cal A} + {\cal B}}{}_{\rm K} \, \Big\}
~~~, \cr 
[ ~ L_{\cal A} \,, \, R_{\cal B}^{{\rm I}_1 \cdots {\rm I}_m}
~ \} &=~ -  \,\, [ ~ {\cal B} \, + \, \fracm 12 \, (m - 2) \, {\cal A}
 ~] \, R_{\cal A + \cal B}^{{\rm I}_1 \cdots {\rm I}_m}
\cr
&{~~~~\,}-~ i \, (i)^{[\fracm {m}2] - [\fracm {m +2}2]} \, [ ~  
\fracm 12 \,{\cal A} \, ({\cal A} + 1  ) ~] \, U_{\cal A + 
\cal B}^{{\rm I}_1 \cdots {\rm I}_m \, J}{}_J  ~~~, }$$
$$\eqalign{
[ ~ G_{\cal A}{}^{\rm I} \,, \, R_{\cal B}^{{\rm J}_1 \cdots {\rm J}_m}
~ \} &=~  2 \,
(i)^{ [\fracm m2 ]  - [\fracm {m+1}2 ]} \,[ ~ {\cal B} \, + \,  
(m - 1) \, {\cal A} \, + \, \fracm 12 \, ~]\, R_{\cal A + \cal 
B}^{{\rm I} \, {\rm J}_1 \cdots {\rm J}_m}
\cr
&{~~~\,}+\,i \,(i)^{ 
[\fracm m2] - [\fracm {m- 1}2 ]} \, \sum_{r=1}^m \, (-1)^{r -1} \, \d^{I \, 
{\rm J}_r} \, R_{\cal A + \cal B}^{{\rm J}_1 \cdots {\rm J}_{r-1} \, 
{\rm J}_{r+1} \cdots {\rm J}_m} \cr 
&{~~~\,}-\,  (-1)^m \, 
(i)^{ [\fracm m2 ] - [\fracm {m + 1}2 ]} \, [~{\cal A} \, +\, \fracm 12 ~]  
\, U_{\cal A  + \cal B}^{{\rm J}_1 \cdots {\rm J}_m \, {\rm I}} \cr
&{~~~\,}+\,i 2  \, 
(i)^{[\fracm {m}2 ] - [\fracm {m + 3}2 ]} \, [~ {\cal A}^2 \, -\, \fracm 14 ~]  
\, U_{\cal A  + \cal B}^{I\, {\rm J}_1 \cdots {\rm J}_m \, K}{}_K ~~~, \cr
[ ~ {\cal D}_{\cal A}{}^{\rm I} \,, \, R_{\cal B}^{{\rm J}_1 \cdots {\rm 
J}_m} ~ \} &=~ i \,2 \,  (i)^{ [\fracm m2 ]  - [\fracm {m+1}2 ]} \,[ ~ {\cal B} 
\, + \,  (m - 1) \, {\cal A} \, + \, \fracm 12 \, ~]\, R_{\cal A + \cal 
B}^{{\rm I} \, {\rm J}_1 \cdots {\rm J}_m}
\cr
&{~~~\,\,}+ \, (i)^{ [\fracm m2] - [\fracm {m- 1}2 ]} \, \sum_{r=1}^m 
\, (-1)^{r -1} \, \d^{{\rm I} \, {\rm J}_r} \, R_{\cal A + \cal B}^{{\rm J}_1 
\cdots {\rm J}_{r-1} \, {\rm J}_{r+1} \cdots {\rm J}_m} \cr 
&{~~~\,\,}+ \, i \,  (-1)^{m}  \, (i)^{ [\fracm m2 ] - [\fracm {m + 1}2 ]}  
\, [~ {\cal A} \, +\, \fracm 12 ~] \,U_{\cal A  + \cal B}^{{\rm J}_1 \cdots 
{\rm J}_m \, {\rm I}} \cr
&{~~~\,\,}-\, 2  \, (i)^{[\fracm {m}2 ] - [\fracm {m + 3}2 ]} \, [~ {\cal 
A}^2 \, -\, \fracm 14 ~] \, U_{\cal A  + \cal B}^{{\rm I}\, {\rm J}_1 \cdots 
{\rm J}_m \, K}{}_K ~~~, \cr
[ ~ R_{\cal A}^{{\rm I}_1 \cdots {\rm I}_m} \,, \, R_{\cal B}^{{\rm J}_1 
\cdots {\rm J}_n} ~ \} 
&=~ -\, (i)^{ [\fracm m2 ] + [\fracm n2 ] - [\fracm {m+n}2 ]} 
 \, [ ~ {\cal A} \, - \, {\cal B} \, + \, \fracm 12 (m \, 
- \, n ) ~] \, R_{\cal A + \cal B}^{{\rm I}_1 \cdots {\rm I}_m \, {\rm J}_1 
\cdots {\rm J}_n} ~~~, \cr
[ ~ G_{\cal A}{}^{\rm I} \, , \, U_{\cal B}^{{\rm J}_1 \cdots {\rm J}_m}
~ \} &=~  2 \, (i)^{ [\fracm m2 ] - [\fracm {m + 1}2 ]} \, [ ~ {\cal B} 
\, + \, (m - 2) \, {\cal A} ~] \, U_{\cal A  + \cal B}^{{\rm I} \, {\rm J}_1 
\cdots {\rm J}_m} \cr
&{~~~\,}- \,2 \, (-1)^{m }  \, (i)^{ [\fracm 
m2 ] - [\fracm {m + 1}2 ]} \, [ ~ {\cal A} \, + \, \fracm 12 ~] \, 
\d^{{\rm I} \, {\rm J}_m} \, U_{\cal A  + \cal B}^{ {\rm J}_1 \cdots 
{\rm J}_{m - 1} K} {}_K \cr
&{~~~\,}+\,i \,  (i)^{ [\fracm m2] - [\fracm {m- 1}2 ]} \, \sum_{r=1}^{m - 1} 
\, (-1)^{r -1} \, \d^{{\rm I} \, {\rm J}_r} \, U_{\cal A + \cal B}^{{\rm J}_1 
\cdots {\rm J}_{r-1} \, {\rm J}_{r+1} \cdots  {\rm J}_m} \cr 
&{~~~\,}-\,i 2 \,  (-1)^{m}  \, (i)^{ [\fracm m2 ] - [\fracm {m - 1}2 ]}\, 
\d^{{\rm I} \, {\rm J}_m} \, R_{\cal A + \cal B}^{{\rm J}_1 \cdots 
{\rm J}_{m-1} } ~~~,~~~, \cr
\left[ ~ {\rm d}_{\cal A}^{\rm I \, \rm J} \, , \, {\rm d}_{\cal 
B}^{\, \rm {K \, L}} \, \right\} 
~=~ &\d^{\rm {I \, K}} \, T_{{\cal A} + {\cal B}}^{\rm J\, \rm L} ~+~
\d^{\rm {I \, L}} \, T_{{\cal A} + {\cal B}}^{\rm J\, \rm K} ~+~
\d^{\rm {J \, K}} \, T_{{\cal A} + {\cal B}}^{\rm I\, \rm L} ~+~
\d^{\rm {J \, L}} \, T_{{\cal A} + {\cal B}}^{\rm I\, \rm K} ~~~,}
$$
$$\eqalign{
[ ~ {\cal D}_{\cal A}{}^{\rm I} \, , \, U_{\cal B}^{{\rm J}_1 \cdots {\rm 
J}_m}~ \} ~=~ & (i)^{ [\fracm m2]  - [\fracm {m- 1}2 ]}  \, \sum_{r=1}^{m 
- 1} \, (-1)^{r -1} \, \d^{{\rm I} \, {\rm I}_r}  \, U_{\cal A + \cal B}^{{\rm 
J}_1 \cdots {\rm J}_{r-1} \, {\rm J}_{r+1} \cdots {\rm J}_m} \cr 
&+\, i \, 2 \, (i)^{ [\fracm m2 ] - [\fracm {m + 1}2 ]} \, [ ~ {\cal 
B} \, + \, (m - 2) \, {\cal A} ~] \, U_{\cal A  + \cal B}^{{\rm I} \, {\rm J}_1 
\cdots {\rm J}_m} \cr
&- \, i \, 2 \,  (i)^{ [\fracm m2 ] - [\fracm {m + 1}2 ]} \, \d^{I \, 
{\rm J}_m} \, [~ {\cal A} \, +\, \fracm 12 ~] \, U_{\cal A  + \cal B}^{{\rm 
I} \, {\rm J}_1 \cdots {\rm J}_{m - 1} K} {}_K \cr
&+\, 2 \,  (-1)^{m}  \, (i)^{ [\fracm m2 ] - [\fracm {m - 1}2 ]}\, 
\d^{{\rm I} \, {\rm J}_m} \, R_{\cal A + \cal B}^{{\rm J}_1 \cdots {\rm 
J}_{m-1}} ~~~, } $$
$$\eqalign{
[ ~ R_{\cal A}^{{\rm I}_1 \cdots {\rm I}_m} \,, \, U_{\cal B}^{{\rm J}_1 
\cdots {\rm J}_n} ~ \} ~=~ &- \, i (-1)^{m n}  \, (i)^{ [\fracm m2 ] + 
[\fracm n2 ] - [\fracm {m+n-2}2 ]} \cr 
& \times \, \sum_{r=1}^m \, (-1)^{r -1} \,\d^{{\rm I}_r \, {\rm J}_n} \, 
R_{\cal A + \cal B}^{{\rm J}_1 \cdots {\rm J}_{n-1} {\rm I}_1 \cdots 
{\rm I}_{r-1}  \, {\rm I}_{r+1} \cdots {\rm I}_m}  ~~~, \cr
&+ \,  (i)^{ [\fracm m2 ] + [\fracm n2 ] - [\fracm {m+n-2}2 ]} \, [ 
~ {\cal B} \, + \, \fracm 12 (m - 2) ~] \, U_{\cal A  + \cal B}^{{\rm I}_1 
\cdots {\rm I}_m  \, {\rm J}_1 \cdots {\rm J}_n } ~~~, \cr
[ ~ U_{\cal A}^{{\rm I}_1 \cdots {\rm I}_m} \,, \, U_{\cal B}^{{\rm J}_1 
\cdots {\rm J}_n} ~ \} ~=~ &i \,  (i)^{ [\fracm m2 ] + [\fracm n2 ] - 
[\fracm {m+n-2}2 ]} \cr
&\times \, \Big\{ \, \sum_{r = 1}^m \, (-1)^{r-1} \, \d^{{\rm I}_m {\rm 
J}_r} \, \, U_{\cal A + \cal B}^{{\rm I}_1 \cdots {\rm I}_{m-1} \, {\rm 
J}_1 \cdots {\rm J}_{r-1} \, {\rm J}_{r+1} \cdots {\rm J}_{n-1} {\rm 
J}_n }~ \cr
&{~~~~} -\, (-1)^{mn} \,  \sum_{r = 1}^m \, (-1)^{r-1} \, \d^{{\rm 
I}_r {\rm J}_n} \, \, U_{\cal A + \cal B}^{{\rm J}_1 \cdots {\rm J}_{
n-1} \, {\rm I}_1 \cdots {\rm I}_{r-1} \, {\rm I}_{r+1} \cdots {\rm 
I}_{m-1} {\rm I}_m } \, \Big\} ~ ~~ , \cr
\left[ ~ {\cal D}_{\cal A}^{\rm I} \, , \, {\rm d}_{\cal B}^{\, \rm {J 
\, K}} \, \right\} 
~=~ &\d^{\rm {I \, J}}  G_{{\cal A} + {\cal B}}^{\, \rm K} ~+~ \d^{\rm {I 
\, K}} G_{{\cal A} + {\cal B}}^{\, \rm J} ~-~ \d^{\rm {J \, K}} G_{{\cal 
A} + {\cal B}}^{\, \rm I} \cr
&{}+\,  2 {\cal B} \, \Big\{ ~  \d^{\rm {I \, J}} U_{{\cal A} + {\cal 
B}}^{\rm {\, K \, L}}{}_{ \, L} ~+~  \d^{\rm {I \, K}} U_{{\cal A} + {\cal 
B} }^{\rm {\, J \, L}}{}_{ \, L} -~ 2 \,\d^{\rm {J \, K}} \, U_{{\cal 
A} + {\cal B}}^{\, \rm{I \, L}}{}_{ \, \rm L} \cr
&{~~~~~~~~~~~}+~ U_{{\cal A} + {\cal B}}^{\, \rm {I \, J \, K}} ~+~ 
U_{{\cal A} + {\cal B}}^{\, \rm {I \, K \, J}}  
~ \Big\}  
~~~.}
\eqno(22) $$

\sect{VI. The Exceptional 1D, $N \,=\, 4$ Super Virasoro D-Module}

The exceptional $N \,=\, 4$ case for the superconformal algebra arose due to 
the possibility of modifying the generic set of generators in (1) and 
replacing a subset of them by the ``deformed'' subset of (4). We can
attempt the same type of approach to the construction of an exceptional
$N \,=\,4$ super Virasoro D-module. This essentially follows the lines of 
the superconformal case. We define $\ell$-deformed vector fields,
$$
\eqalign{ 
G_{\cal A} {}^{\rm I} ~\equiv~ i \, &\t^{{\cal A} + \fracm 12}  \, \Big[\, 
\, \pa^{\rm I} ~-~ i \, 2 \, \zeta^{\rm I} \, \pa_{\t} ~ \Big] ~+~ 2 (\, 
{\cal A} \,+\, \fracm 12 \,) \t^{{\cal A} - \fracm 12}  \z^{\rm I} \z^{\rm 
K} \, \pa_{\rm K} {~~~~~~~~~~~~~~~~~~~} \cr
&+\,   \ell \, (\, {\cal A} + \fracm 12 \,) \,  \t^{{\cal A} - \fracm 
12} \, \Big[\, \e^{\rm I \, \rm J \, \rm K \, \rm L }
 \, \z_{\rm J} \z_{\rm K} \, \pa_{\rm L} 
~-~ i \, 4  \, \zeta^{(3) \, \rm I} \, \pa_{\t} ~ \Big]  \cr
&+~ i \, 4 \, \ell\, (\, {\cal A}^2 - \fracm 14 \,) \,  
\t^{{\cal A} - \fracm 32}  \, \z^{(4)} \, \pa^{\rm I}  ~~~, }
$$ 
$$
\eqalign{ {~~~~~~~~}
L_{\cal A} &\equiv~ -\, \Big[ \, \t^{{\cal A} + 1} \pa_{\t} ~+~ \fracm 12
({\cal A} \, + \, 1) \, \t^{\cal A} \zeta^{\rm I} \, \pa_{\rm I} ~ \Big] 
~~~ \cr
&~~~~\, +\, i \,  \ell \, {\cal A} \, ({\cal A} + 1) \,  \t^{{\cal A} - 1} 
\Big[~  \zeta^{(3) \, \rm I} \, \pa_{\rm I}  \, + \, i\, 4  \z^{(4)}
 \pa_{\t} ~ \Big]  ~~~, \cr
T_{\cal A}^{\, \rm {I \, J }} &\equiv~ \t^{\cal A} \, \Big[ ~ \z^{\rm I} 
\, \pa^{\rm J} ~-~ \z^{\rm J} \, \pa^{\rm I} ~-~ \ell \, \e^{\rm I \, 
\rm J \, \rm K \, \rm L } \z_{\rm K} \, \pa_{\rm L}
 ~\Big] \cr
&~~~~-~ i \, 2 \, \ell \, {\cal A} \, \t^{{\cal A} - 1} \, \Big[ ~  \z^{(3) \, 
\rm I} \, \pa^{\rm J} ~-~ \z^{(3) \, \rm J} \, \pa^{\rm I} ~-~ 
 \ell \, \e^{\rm I \, \rm J \, \rm K \, \rm L } \z^{(3)}{}_{\rm K} 
\, \pa_{\rm L}  ~\Big] ~~~.}
\eqno(23) $$
It is a matter of simple but long and intricate calculations to show 
that these derivations provide a representation of an $N \,=\,4$ super 
Virasoro algebra.  We find the closure of the $\ell$-dependent set $\{ 
L_{\cal A} , \, G_{\cal A} {}^{\rm I} , \, T_{\cal A}^{\, \rm {I \, J }} 
\}$ under commutation compelling given the number of identities required 
to achieve it. It should be clear from our method that {\it {only}} in 
the extended cases of $N \,=\, 2$ and $4$, does the set $\{ L_{\cal A} , \, 
G_{\cal A} {}^{\rm I} , \, T_{\cal A}^{\, \rm {I \, J }} \}$ close. 
For all other extended values of $N$, additional operators (contained 
in the super covering ${\cal G}{\cal R}$ D-module) are required for 
closure. Another implication of our construction is that there exists 
a distinct ``large'' $N \, =\,4$ super ${\cal G} {\cal R}$ Virasoro algebra 
that appears as a subset of the operators in (21). 

\sect{VII. Conclusions}

Since the super ${\cal G}{\cal R}$ covering algebra is embedded as an
algebra of derivations, it is a geometrical property of 1D, $N$-extended 
superspace itself and does not depend on the existence of any dynamical 
quantities. Alternately, we can now undertake investigations of 
arbitrarily extended super Virasoro algebras using super differential 
operators and super functions as opposed to the traditional use of 
Hilbert spaces and objects contained therein.  It is clear (by the 
same argument that we used to oxidize the 1D, $N$-extended superconformal 
algebra to 2D (1,0) superspace) that there exist super 2D ($p,\, q$) 
${\cal G}{\cal R}$ covering algebras. Our approach shows how the 2D 
generators of super Virasoro algebras can be constructed from the 
same vector fields used for 1D theories by following the oxidation 
procedure described below (4) in section two.  In any event, our approach 
offers a field-independent way to study arbitrary $N$-extensions of
super-Virasoro algebras in 1D superspace as well as arbitrary 
(p, q)-extensions in 2D superspace. 

The behavior of $T_{\cal A}^{\, \rm {I \, J }}$ as the value of
$N$ changes is very characteristic and reminiscent of another system
we have previously investigated \cite{GRn3}.  There it was shown
that there exist matrices denoted by $f^{\rm {I \, J }}$ which occur
in the off-shell description of $N$-extended NSR spinning particles. 
For all values of $N$, except four, these matrices were representations
of O($N$). For the case of $N \,=\,4$, these were representations of SU(2).
We are thus led to suspect that the $f^{\rm {I \, J }}$'s provide
a matrix representation of the $T_{0}^{\, \rm {I \, J }}$ generators.
If this suggestion is accepted, then the vector spaces upon which the 
$f^{\rm {I \, J }}$-matrices act can be interpreted as the spinor 
representations of the super Virasoro algebra. The ${\cal GR}$(d, $N$)
algebras discussed in ref. \cite{GRn3} would then be related to the 
super Virasoro algebra in exactly the same way that the Spin(d) algebra 
is related to the conformal algebra. With this new interpretation, we 
may say that the $\aleph_0$-extended off-shell NSR spinning particle
models we constructed required the use of ``Virasoro spinor representations.''

As we noted, the super ${\cal G}{\cal R}$ covering algebra is larger
than the super Virasoro algebra. Here the most interesting operator 
seems to be ${\cal D}_{\cal A} {}^{\rm I}$. It bares a striking similarity
(even in form) to the covariant Green-Schwarz superstring operator 
$D_{\a}(\s)$, first introduced by Siegel \cite{SGL}. In turn the zero 
mode of $D_{\a}(\s)$ corresponds to the usual superspace ``supercovariant 
derivative.'' We still have before us the task of fully understanding 
the significance of ${\cal D}_{\cal A} {}^{\rm I}$. The most naive 
expectation is that it is an NSR analog to $D_{\a}(\s)$.

Other interesting future avenues to study are the behaviors of the 
realizations of the super Virasoro algebra that are induced by our 
vector fields on restricted super functions of the form
$$
{\cal F}(\t, \, \z^{\rm I}) ~=~ \sum_{\cal A} \, \, \t^{\cal A}\, a_{(\cal A)} 
(\z^{\rm I}) ~~~, 
\eqno(24) $$
in the 1D case and 
$$
{\cal F}(\t, \, \s , \, \z^{+ \, {\rm I}_R} , \, \z^{- \, {\rm I}_L}
) ~=~ \sum_{{\cal A}, \,{\cal B}} \, \, (\s^{\pp})^{\cal A} \, (\s^{\mm})^{\cal B} \, 
a_{(\cal A , \, \cal B) }  (\z^{+ \, {\rm I}_R} , \, \z^{- \, 
{\rm I}_L} )  ~~~, 
\eqno(25) $$
in the 2D case\footnote{It is a simple step to append a space-time
index to these in order to introduce quantities \newline ${~~~~\,}$ 
that bare a resemblance to particles and strings ${\cal F} \to X^{\underline 
m}$.}.  Clearly such restricted super functions form a closed set under 
the operators in (21) as well as ordinary multiplication.

Looking back at the $N$ = 0 truncation of our results, we see that there
remains only the operator $L_{\cal A}$. This again raises the logical
question of the import of the operator $H_r$? It remains present in 
the purely bosonic limit.  We have successfully reached our goal of a 
model-independent representation of the super Virasoro algebras for all 
values of $N$. An important next step is to investigate whether any of 
the new purely ``kinematical'' features (e.g. $H_r$, ${\cal D}_{\cal A}$, 
etc.) we have proposed yield new insights into dynamical systems such 
as $N$-extended supergravity as well as super and heterotic strings.  
Apothegmatically, do our model-independent reformulations of super 
Virasoro algebras matter?
$${~~~}$$
``{\it {There is geometry in the humming of the strings.}}'' -- Pythagoras
$${~~~}$$
\noindent
{\bf {Acknowledgement}} \newline \noindent
${~~~~}$The authors wish to acknowledge discussions with 
T.~H\" ubsch and W. Siegel.  


\end{document}